%% file: main.tex
\documentclass[11pt]{article}

\usepackage{amsmath, amssymb, amsthm}
\usepackage{hyperref}

\usepackage{graphicx} 
\usepackage{float}    
\usepackage{tikz}
\usetikzlibrary{arrows.meta, positioning, shapes.misc, fit, backgrounds}

\newtheorem{remark}{Remark}

\title{A Clarifying Note on Long-Horizon Investment and Dollar-Cost Averaging:\\
An Effective Investment Exposure Perspective}
\author{Zeusu Sato}
\date{}

\begin{document}
\maketitle

\begin{abstract}
It is widely claimed in investment education and practice that extending
the investment horizon reduces risk, and that diversifying investment
timing---for example through dollar-cost averaging (DCA)---further
mitigates investment risk.
Although such claims are intuitively appealing, they are often stated
without precise definitions of risk or a clear separation between risk
and uncertainty.

This paper revisits these two beliefs within a unified probabilistic
framework.
We define risk at the expectation level as a property of the generating
distribution of cumulative investment outcomes, and distinguish it from
uncertainty, understood as the dispersion of realized outcomes across
possible paths.
To enable meaningful comparisons across horizons and investment
schedules, we introduce the notion of effective investment exposure,
defined as time-integrated invested capital.

Under stationary return processes with finite variance, we show that
extending the investment horizon does not alter expected risk, expected
return, or the risk--return ratio on a per-unit-exposure basis.
In contrast, different investment timing strategies can induce distinct
exposure profiles over time.
As a result, lump-sum investment and dollar-cost averaging may differ not
only in uncertainty but also in expected risk when compared at equal
return exposure, although the resulting risk differences are of constant
order and do not grow with the investment horizon.

We further show that both horizon and timing comparisons generate
systematic differences at the level of uncertainty.
Longer horizons lead to greater concentration of realized outcomes than
shorter horizons, while uniform temporal exposure leads to greater
concentration than time-diversified schedules such as DCA.
These effects, however, concern uncertainty rather than risk itself.

Taken together, the results clarify why common narratives surrounding
long-horizon investment and dollar-cost averaging are conceptually
misleading, while also explaining why adopting such strategies under
budgetary or timing constraints need not be regarded as irrational.
\end{abstract}

\input{sections/01_introduction}
\input{sections/02_setup_and_notation}
\input{sections/03_long_term_investment}
\input{sections/04_timing_diversification}
\input{sections/05_discussion}
\input{sections/06_conclusion}

\appendix
\input{sections/99a_appendix}
\input{sections/99b_appendix}
\input{sections/99c_appendix}
\input{sections/99d_appendix}

\bibliographystyle{plain}
\bibliography{refs}

\end{document}

%% file: sections/01_introduction.tex
\section{Introduction}

In discussions of long-term investment and portfolio construction, it is often
claimed that extending the investment horizon reduces risk, or that
diversifying investment timing---for example through dollar-cost
averaging---reduces investment risk.
Such claims are widely encountered in financial literacy education and in the
sales and marketing of financial products, yet they are rarely stated with
precise mathematical formulations.

The purpose of this note is to clarify two common misconceptions.
The first concerns the belief that investment risk decreases as the holding
period becomes longer.
The second concerns the belief that diversification in investment timing,
exemplified by dollar-cost averaging, reduces investment risk.
Although these two claims are often discussed together, they involve distinct
mechanisms and must be analyzed at different conceptual levels.
In this paper, the two issues are treated in parallel within a unified
probabilistic framework.

A key feature of our approach is an explicit separation between
expectation-level quantities (such as expected return and expected risk) and
the dispersion of realized outcomes, which we refer to as uncertainty.
We further emphasize the role of exposure normalization in making meaningful
comparisons across investment horizons and timing strategies.
This framework allows us to distinguish effects that arise from changes in the
generating distribution from those that arise solely from finite-sample
variation.

A central motivation for this approach is reproducibility.
Numerical simulations can yield qualitatively different conclusions depending
on parameter choices, while empirical analyses may depend strongly on the
sampling period, market, or region under consideration.
By contrast, mathematical analysis makes it possible to isolate statements that
hold universally, independent of when, where, or on which market the analysis
is performed.
The results presented here are therefore statements of necessity, rather than
observations contingent on specific data.

Some of the individual ideas discussed in this paper are classical.
Related arguments can be traced back to Samuelson’s critique of time
diversification and to subsequent analyses of dollar-cost averaging.
However, these ideas are often presented only partially or in isolation.
As a result, their logical relationships---and the precise conditions under
which they do or do not apply---are frequently obscured.
The present note aims to restate and unify these arguments in a concise and
self-contained manner, while providing explicit quantitative comparisons where
appropriate.

This note does not advocate or discourage any particular investment strategy.
Its sole objective is to clarify how risk and uncertainty, defined in terms of
the distribution of cumulative investment outcomes, behave under changes in
the investment horizon and investment timing.
We show that extending the investment horizon does not alter expected risk or
expected return on a per-unit-exposure basis, although it does reduce
uncertainty.
In contrast, we show that different timing strategies, such as lump-sum
investment and dollar-cost averaging, can induce distinct exposure profiles
that lead to quantitative differences in expected risk when compared at equal
return exposure, even though these differences remain of constant order.

The remainder of the paper is organized as follows.
Section~2 introduces the basic setup and notation.
Section~3 analyzes the scaling of cumulative risk with the investment horizon
and shows that extending or shortening the horizon does not create an
expectation-level advantage once exposure is normalized.
Section~4 examines diversification in investment timing and demonstrates that,
while dollar-cost averaging affects uncertainty, it is not optimal in the
sense of minimizing expected risk per unit exposure.
Section~5 discusses model simplicity and explains why the main conclusions
remain unchanged unless extreme or unrealistic assumptions are introduced.
Appendix~A and Appendix~D provide additional technical details and clarify the
relationship between time-series and cross-sectional notions of risk.

%% file: sections/02_setup_and_notation.tex
\section{Setup and Notation}

\subsection{Returns as a stationary data-generating process}

We consider a discrete-time return process $\{R_t\}_{t\ge1}$, modeled
throughout as log-returns.
If $P_t$ denotes the asset price at time $t$, we define
\begin{equation}
R_t := \log P_t - \log P_{t-1}.
\end{equation}

The return process is assumed to be stationary, so that
\[
\mathbb{E}[R_t] = \mu,
\qquad
\mathrm{Var}(R_t) = \sigma^2
\]
for all $t$.
No assumption on independence is imposed at this stage unless explicitly
stated.

\subsection{Pathwise cumulative quantities}

For a horizon of length $t$, we define the cumulative (pathwise) return as the
random variable
\begin{equation}
\mathrm{Sum}(t) := U_t := \sum_{i=1}^{t} R_i .
\end{equation}

We also define the cumulative squared deviation by
\begin{equation}
\mathrm{Dev}^2(t) := V_t := \sum_{i=1}^{t} (R_i-\mu)^2 .
\end{equation}

Both quantities are random variables constructed from the same return
sequence $(R_1,\dots,R_t)$.
At this stage, they are purely pathwise objects and do not yet constitute
notions of return, risk, or uncertainty.

\subsection{Return and risk as expectation-level quantities}

We define \emph{Return} and \emph{Risk} as expectation-level quantities derived
from the pathwise objects above:
\begin{equation}
\mathrm{Return}(t) := \mathbb{E}[\mathrm{Sum}(t)],
\qquad
\mathrm{Risk}(t) := \mathbb{E}[\mathrm{Dev}^2(t)] .
\end{equation}

Under the assumption that $\{R_t\}$ are independent and identically distributed,
linearity of expectation yields
\begin{equation}
\mathrm{Return}(t) = t\mu,
\qquad
\mathrm{Risk}(t) = t\sigma^2 .
\end{equation}

In the Gaussian case, one further has
\[
\mathrm{Sum}(t) \sim \mathcal{N}\bigl(\mathrm{Return}(t),\,\mathrm{Risk}(t)\bigr),
\qquad
\frac{\mathrm{Dev}^2(t)}{\sigma^2} \sim \chi^2_t .
\]
Thus, Return and Risk coincide with the location and scale parameters of the
generating distribution of cumulative returns.
In this sense, Risk is not an arbitrary summary statistic, but a structural
parameter of the data-generating process itself.

\subsection{Uncertainty as dispersion across realizations}

While Return and Risk are defined as expectations, realized outcomes vary across
possible paths.
We quantify this dispersion by defining \emph{uncertainty} as the variance of
the corresponding pathwise quantities:
\begin{equation}
\mathrm{Var}[\mathrm{Sum}(t)],
\qquad
\mathrm{Var}[\mathrm{Dev}^2(t)] .
\end{equation}

Under independence and finite moments,
\begin{equation}
\mathrm{Var}[\mathrm{Sum}(t)] = t\,\sigma^2,
\end{equation}
and, writing $Y_i := (R_i-\mu)^2$,
\begin{equation}
\mathrm{Var}[\mathrm{Dev}^2(t)]
= t\,\mathrm{Var}(Y_1)
= t\!\left(
\mathbb{E}\!\left[(R_1-\mu)^4\right] - \sigma^4
\right),
\end{equation}
provided the fourth moment of $R_t$ exists.

Risk and uncertainty are therefore conceptually distinct.
Risk characterizes the scale of the generating distribution, while uncertainty
describes how widely realized quantities may vary across paths.

The distinction between expectation-level quantities and sample variability
is standard in probability and statistics; see, e.g.,
Billingsley (1995) \cite{billingsley1995} and Wasserman (2004) \cite{wasserman2004}.

\subsection{Observed paths and an ensemble interpretation}

In practice, only a single realized path of returns
\[
r_1, r_2, \dots, r_t
\]
is observed.
From this realization, we define the realized cumulative return
\begin{equation}
u_t := \sum_{i=1}^{t} r_i
\end{equation}
and the realized sample mean
\begin{equation}
m_t := \frac{1}{t}\sum_{i=1}^{t} r_i .
\end{equation}

Since the population mean $\mu$ is unknown, we define the realized cumulative
squared deviation by
\begin{equation}
v_t := \sum_{i=1}^{t} (r_i - m_t)^2 ,
\end{equation}
which corresponds to $(t-1)$ times the realized sample variance.

Expectations and variances of such sample-based quantities are interpreted with
respect to an ensemble of return paths
\[
\{r^{(i)}_1,\dots,r^{(i)}_t\}_{i=1}^N,
\]
each generated by the same stationary process.
This ensemble interpretation, standard in empirical finance, aligns naturally
with a Bayesian viewpoint.

For example, defining
\[
\bar U_{t,N} := \frac{1}{N}\sum_{i=1}^N U_t^{(i)},
\]
one has
\begin{equation}
\mathbb{E}[\bar U_{t,N}] = t\mu,
\qquad
\mathrm{Var}(\bar U_{t,N}) = \frac{t\sigma^2}{N}.
\end{equation}
An analogous relation holds for cumulative risk.
Thus, while expected quantities depend only on the horizon $t$, uncertainty
depends jointly on $t$ and the number of independent paths $N$.

The ensemble interpretation adopted here is standard and aligns naturally
with a Bayesian viewpoint; see, e.g., Gelman et al. (2013) \cite{gelman2013}.

\subsection{Effective investment exposure}

When comparing investment strategies, it is essential to distinguish between
nominal invested capital and the time profile of exposure.
To formalize this distinction, we introduce the notion of
\emph{effective investment exposure}.

Let $w_i\in[0,1]$ denote the fraction of total capital invested at time $i$, with
$\sum_{i=1}^t w_i = 1$.
The effective investment exposure over a horizon of length $t$ is defined as
\[
E := \sum_{i=1}^t w_i (t-i+1),
\]
representing time-integrated invested capital.

Although different investment schedules may result in the same total invested
capital, their effective exposures can differ substantially.
Comparisons of cumulative return or cumulative risk that do not normalize by
exposure therefore conflate fundamentally different objects.

This notion will play a central role in Sections~3 and~4, where it clarifies why
comparisons across horizons or timing strategies are uninformative without
appropriate normalization.

\subsection{Scope of the analysis}

The objective of this paper is to analyze the probabilistic structure of return,
risk, and uncertainty.
Accordingly, no preference specification, behavioral assumption, or optimization
criterion is introduced.
The comparisons considered in this paper---long versus short horizons and
lump-sum versus time-diversified investment---do not alter expected cumulative
return or expected cumulative risk under the assumptions of this section.

The only systematic differences that arise concern uncertainty, defined as the
dispersion of realized outcomes.
Distinguishing clearly between pathwise quantities, expectation-level risk, and
uncertainty is essential for avoiding common misinterpretations.
Throughout the paper, all expectations and variances are taken with respect to
the stationary data-generating process introduced above.

%% file: sections/03_long_term_investment.tex
\section{Misconception I: ``Long-Term Investment Reduces Risk''}

The fallacy of time diversification has long been recognized,
dating back at least to Samuelson (1994) \cite{samuelson1994}.

A common belief in investment education and practice is that extending the
investment horizon inherently reduces risk.
In this section, we show that this belief is false once the hierarchy of
objects in Section~2 is kept explicit.
Longer horizons may reduce \emph{uncertainty} (dispersion of realized outcomes),
but they do not reduce \emph{risk}, understood as an expectation-level
parameter of the generating distribution.

\subsection{Pathwise quantities versus expectation-level risk}

Assume that returns $\{R_t\}$ are independent and identically distributed with
mean $\mu$ and variance $\sigma^2$.
Recall from Section~2 that the pathwise cumulative quantities are
\[
\mathrm{Sum}(t) := U_t=\sum_{i=1}^t R_i,
\qquad
\mathrm{Dev}^2(t) := V_t=\sum_{i=1}^t (R_i-\mu)^2,
\]
and that \emph{Return} and \emph{Risk} are defined at the expectation level by
\[
\mathrm{Return}(t):=\mathbb{E}[U_t],
\qquad
\mathrm{Risk}(t):=\mathbb{E}[V_t].
\]

By linearity of expectation,
\begin{equation}
\mathrm{Return}(t)=t\mu,
\qquad
\mathrm{Risk}(t)=t\sigma^2.
\end{equation}
Thus, risk increases linearly with the horizon length $t$.
There is no sense in which risk decreases as the investment horizon grows.

In the Gaussian case, one further has
\[
U_t \sim \mathcal{N}\bigl(\mathrm{Return}(t),\,\mathrm{Risk}(t)\bigr),
\qquad
\frac{V_t}{\sigma^2} \sim \chi^2_t,
\]
which makes explicit that the generating distribution does not become
``less risky'' as $t$ increases.

\begin{remark}[On time scaling and a common misconception]
The scaling with $t$ in $\mathrm{Risk}(t)=t\sigma^2$ is for a variance-type
quantity, not for a standard deviation.
In particular, $\mathrm{Var}(U_t)=t\sigma^2$ and $\mathrm{SD}(U_t)=\sqrt{t}\,\sigma$.
Therefore, statements of the form ``risk scales as $1/t$'' obtained by dividing
a standard deviation by $t$ are merely scaling errors.
In practice, time normalization of volatility is performed by dividing by
$\sqrt{t}$.
\end{remark}

\subsection{Short-term vs long-term: an exposure-normalized perspective}

The preceding subsection shows that comparing horizons using unnormalized
cumulative risk is uninformative: shorter horizons trivially exhibit smaller
cumulative risk simply because they involve less exposure.
This triviality is precisely what effective investment exposure in
Section~2 captures.

Meaningful comparisons must be made after normalizing by exposure.
In the present setting, the natural per-unit-time normalization corresponds to
exposure normalization, and we define the annualized expected return and risk by
\begin{equation}
\mathrm{A.Return} := \frac{\mathrm{Return}(t)}{t} = \mu,
\qquad
\mathrm{A.Risk} := \sqrt{\frac{\mathrm{Risk}(t)}{t}} = \sigma.
\end{equation}
The corresponding risk--return ratio is then
\begin{equation}
\mathrm{RR} := \frac{\mathrm{A.Return}}{\mathrm{A.Risk}} = \frac{\mu}{\sigma},
\end{equation}
which is invariant with respect to the horizon length $t$.
From the perspective of exposure-normalized risk--return trade-offs, neither
short-term nor long-term investment admits an intrinsic advantage.

The only systematic difference between short and long horizons concerns
uncertainty, which we now analyze explicitly.
As shown in Section~2, the dispersion of realized per-unit-time quantities,
such as $U_t/t$ or $V_t/t$, decreases with $t$.
This reduction reflects concentration of realized outcomes, not a reduction in
risk itself.
Confusing these two notions is the source of the misconception that long-term
investment ``reduces risk.''

\begin{remark}[Relation to $t$-statistics]
If one were to define a risk--return ratio using cumulative quantities,
$\mathbb{E}[U_t]/\sqrt{\mathbb{E}[V_t]}$, the resulting expression would scale
as $\sqrt{t}$.
This behavior is identical to the $\sqrt{N}$ factor in classical $t$-statistics.
Such scaling does not indicate that the mean return $\mu$ increases or that the
per-period risk $\sigma$ decreases.
Rather, it reflects the fact that statistical evidence for a nonzero mean
accumulates with repeated sampling.
\end{remark}

\begin{remark}[Relation to utility-based preferences]
In economic theory, comparing risk--return trade-offs is often formalized by
introducing a utility function and assuming that investors are risk-averse.
The present analysis does not rely on any such specification.
Indeed, the very act of comparing risk magnitudes already presumes that the
decision maker is not risk-neutral in the economic sense.
More generally, the result obtained here does not depend on whether investors
are risk-averse or risk-seeking.
As long as preferences are monotone in expected return and depend on risk only
through its magnitude, neither short-term nor long-term investment admits an
intrinsic optimum. Only when utility is nonlinear in risk itself does the
choice of horizon become preference-dependent.
\end{remark}

\subsection{Uncertainty of realized outcomes}

We next formalize the statement that shorter horizons exhibit greater
``uncertainty'' of realized outcomes.
Here uncertainty refers to the dispersion of the pathwise quantities
$\mathrm{Sum}(t)=U_t$ and $\mathrm{Dev}^2(t)=V_t$ across possible realizations,
i.e.\ $\mathrm{Var}(U_t)$ and $\mathrm{Var}(V_t)$, and is conceptually distinct
from the expectation-level risk $\mathrm{Risk}(t)=\mathbb{E}[V_t]$.

\paragraph{Uncertainty of cumulative return.}
Under i.i.d.\ returns,
\begin{equation}
\mathrm{Var}(U_t) = t\,\sigma^2.
\label{eq:Var_Ut}
\end{equation}
Consequently,
\begin{equation}
\mathrm{Var}\!\left(\frac{U_t}{t}\right)
= \frac{\sigma^2}{t},
\label{eq:Var_Ut_over_t}
\end{equation}
which decreases as $t$ increases, expressing concentration of realized average
returns around $\mu$.

\paragraph{Uncertainty of cumulative squared deviation.}
Define $Y_i := (R_i-\mu)^2$ so that $V_t=\sum_{i=1}^t Y_i$ and
$\mathbb{E}[Y_i]=\sigma^2$.
If $\mathrm{Var}(Y_1)<\infty$ (equivalently, $R_i$ has a finite fourth moment),
then
\begin{equation}
\mathrm{Var}(V_t) = t\,\mathrm{Var}(Y_1),
\label{eq:Var_Vt}
\end{equation}
and hence
\begin{equation}
\mathrm{Var}\!\left(\frac{V_t}{t}\right)
= \frac{\mathrm{Var}(Y_1)}{t}
\;\longrightarrow\; 0
\qquad (t\to\infty).
\label{eq:Var_Vt_over_t}
\end{equation}
Thus, while expected risk $\mathrm{Risk}(t)=t\sigma^2$ grows linearly with $t$,
the realized per-unit-time statistic $V_t/t$ becomes increasingly concentrated
around $\sigma^2$ as the horizon increases.

Equations \eqref{eq:Var_Ut_over_t} and \eqref{eq:Var_Vt_over_t} provide a precise
sense in which shorter horizons are more likely to produce tail realizations
(both positive and negative) than longer horizons, even under a stationary
data-generating process.
Importantly, this phenomenon reflects concentration of realized outcomes and
should not be interpreted as a change in expected return or in population-level
risk per unit time.

%% file: sections/04_timing_diversification.tex
\section{Misconception II: ``Timing Diversification Reduces Risk''}

Standard investment texts note that dollar-cost averaging does not provide a
general dominance relation over lump-sum investment; see, e.g., Bodie, Kane,
and Marcus (2018) \cite{bodie2018}.

A second common belief in investment practice is that diversifying investment
timing---most notably through dollar-cost averaging (DCA)---reduces investment
risk relative to lump-sum investment.
In this section, we show that this belief rests on the same conceptual error
as the long-horizon fallacy discussed in Section~3.
Once the hierarchy of objects in Section~2 is kept explicit, the comparison
splits naturally into (i) expectation-level quantities (Return and Risk) and
(ii) uncertainty (dispersion across paths).
The key point is that timing diversification changes exposure profiles and
thereby affects dispersion, but it does not create an intrinsic risk advantage.

\subsection{Pathwise timing strategies}

We consider a fixed investment budget, normalized to one unit of capital, and a
finite horizon of length $t$.
Let $\{R_i\}_{i=1}^t$ denote the return process introduced in Section~2.

\paragraph{Lump-sum investment.}
All capital is invested at the initial time $i=1$.
The pathwise cumulative return and squared deviation are
\[
U_t^{\mathrm{LS}} := \sum_{i=1}^t R_i,
\qquad
V_t^{\mathrm{LS}} := \sum_{i=1}^t (R_i-\mu)^2.
\]

\paragraph{Time-diversified investment (DCA).}
Let $w_i\in[0,1]$ denote the fraction of capital invested at time $i$, with
$\sum_{i=1}^t w_i=1$.
Define the cumulative invested fraction at time $j$ by
\[
a_j := \sum_{i=1}^j w_i,
\qquad 0\le a_j\le 1.
\]
Then the pathwise cumulative return can be written as
\[
U_t^{\mathrm{TD}}
:= \sum_{i=1}^t w_i \sum_{j=i}^t R_j
= \sum_{j=1}^t a_j R_j,
\]
and the corresponding pathwise squared-deviation statistic is
\[
V_t^{\mathrm{TD}}
:= \sum_{j=1}^t (a_j R_j - a_j\mu)^2
= \sum_{j=1}^t a_j^2 (R_j-\mu)^2.
\]
These are pathwise objects, analogous to $\mathrm{Sum}(t)$ and $\mathrm{Dev}^2(t)$
in Section~2.

\subsection{Effective exposure and an extreme example}

The key difference between lump-sum and time-diversified investment lies not in
timing diversification itself, but in the exposure profile over time.

To see this most clearly, consider an extreme case:
suppose no capital is invested until the final period $i=t$, at which point the
entire budget is invested.
The terminal invested amount is identical to that of lump-sum investment, yet
exposure over most of the horizon is zero.
It is therefore evident that cumulative quantities constructed from exposure
over time must differ across timing strategies, even when the final invested
amount is the same.

This observation motivates normalizing comparisons by exposure measures, as in
Section~2.

\subsection{Return and risk at the expectation level}

We now compare lump-sum and time-diversified investment at the expectation
level.
By linearity of expectation,
\[
\mathbb{E}[U_t^{\mathrm{LS}}] = \mu \sum_{j=1}^t 1 = t\mu,
\qquad
\mathbb{E}[U_t^{\mathrm{TD}}] = \mu \sum_{j=1}^t a_j .
\]
Thus, expected cumulative return scales with the \emph{return exposure}
\[
E^{(1)} := \sum_{j=1}^t a_j,
\]
which equals $t$ for lump-sum investment and $(t+1)/2$ for evenly spaced DCA.

For the squared-deviation statistic,
\[
\mathbb{E}[V_t^{\mathrm{LS}}] = \sigma^2 \sum_{j=1}^t 1 = t\sigma^2,
\qquad
\mathbb{E}[V_t^{\mathrm{TD}}] = \sigma^2 \sum_{j=1}^t a_j^2 .
\]
Thus, expected cumulative risk scales with the \emph{risk exposure}
\[
E^{(2)} := \sum_{j=1}^t a_j^2,
\]
reflecting the variance-type (quadratic) structure of risk accumulation.

Normalizing by these exposures yields the exposure-normalized quantities
\[
\widetilde{\mathrm{Return}}
:= \frac{\mathbb{E}[U_t]}{E^{(1)}} = \mu,
\qquad
\widetilde{\mathrm{Risk}}
:= \frac{\mathbb{E}[V_t]}{E^{(2)}} = \sigma^2.
\]
Hence, timing diversification does not improve expected return per unit
\emph{return exposure} nor expected risk per unit \emph{risk exposure};
it only reshapes exposure profiles.

\subsection{Uncertainty of realized outcomes}

We next examine uncertainty, defined as dispersion of realized quantities
around their expectations.
Unlike the horizon comparison in Section~3, the direction of this effect can be
counterintuitive in the present setting.

\paragraph{Uncertainty of cumulative return.}
Under independence,
\[
\mathrm{Var}(U_t^{\mathrm{TD}})
= \mathrm{Var}\!\left(\sum_{j=1}^t a_j R_j\right)
= \sigma^2 \sum_{j=1}^t a_j^2 .
\]
For lump-sum investment, $a_j\equiv 1$ and $\mathrm{Var}(U_t^{\mathrm{LS}})=t\sigma^2$.
For evenly spaced DCA, $a_j=j/t$, yielding
\[
\mathrm{Var}(U_t^{\mathrm{DCA}})
= \sigma^2 \sum_{j=1}^t \left(\frac{j}{t}\right)^2
= \sigma^2 \frac{t(t+1)(2t+1)}{6t^2}
\sim \frac{t}{3}\sigma^2
\qquad (t\to\infty).
\]
Thus, the variance of cumulative return is smaller under time-diversified
investment, reflecting reduced exposure over most of the horizon.

\paragraph{Uncertainty per unit exposure.}
To compare dispersion on an equal footing, we normalize by return exposure
$E^{(1)}=\sum_{j=1}^t a_j$ and define
\[
\tilde U_t := \frac{U_t}{E^{(1)}}.
\]
Then
\[
\mathrm{Var}(\tilde U_t)
= \frac{\sigma^2 \sum_{j=1}^t a_j^2}{\left(\sum_{j=1}^t a_j\right)^2}.
\]
For lump-sum investment,
\[
\mathrm{Var}(\tilde U_t^{\mathrm{LS}})=\frac{\sigma^2}{t},
\]
while for evenly spaced DCA,
\[
\mathrm{Var}(\tilde U_t^{\mathrm{DCA}})
= \sigma^2
\frac{\sum_{j=1}^t (j/t)^2}{\left(\sum_{j=1}^t j/t\right)^2}
\sim \frac{4\sigma^2}{3t}
\qquad (t\to\infty).
\]
Hence, even after exposure normalization, lump-sum investment exhibits
\emph{smaller uncertainty} of realized per-unit-exposure outcomes than
time-diversified investment.

\paragraph{Interpretation.}
This result may appear counterintuitive, as dollar-cost averaging is often
described as a form of ``diversification.''
However, the diversification here operates across \emph{time}, not across
independent risk sources.
Capital invested at different times shares overlapping return components and is
therefore perfectly positively correlated over those overlaps.
As a result, temporal averaging is weaker under time-diversified schedules,
leading to greater dispersion of realized per-unit-exposure outcomes.

\subsection{Structural equivalence with the horizon fallacy}

The misconception that timing diversification reduces risk is structurally
identical to the belief that long-term investment reduces risk.
In both cases, apparent improvements arise from conflating changes in exposure
or dispersion with changes in risk itself.

Once comparisons are normalized by the appropriate exposure measures, expected return per unit \emph{return exposure} and expected risk per unit \emph{risk exposure} are invariant.
The only residual distinction concerns uncertainty (dispersion of realized
outcomes), which reflects concentration effects rather than changes in risk.
Recognizing this common structure clarifies why neither investment horizon nor
investment timing admits an intrinsic risk advantage.

\subsection{Cross-sectional and temporal diversification: a unified view}

Here we use variance only to highlight the classical diversification mechanism; the time-series notion of risk used in this paper remains expectation-based as defined in Section~2.
The analysis above concerns changes in the timing of exposure to a single risk
source.
To fully clarify the role of diversification, it is useful to place timing
diversification alongside classical cross-sectional diversification within a
common mathematical framework.

\paragraph{A general diversification template.}
Let $\{R_\alpha\}_{\alpha\in\mathcal{I}}$ denote a collection of risk sources,
and let $\{a_\alpha\}_{\alpha\in\mathcal{I}}$ be nonnegative exposure weights.
The aggregate return is
\[
U := \sum_{\alpha\in\mathcal{I}} a_\alpha R_\alpha,
\]
and meaningful comparisons require normalization by total exposure,
\[
\sum_{\alpha\in\mathcal{I}} a_\alpha = A \quad \text{(fixed)}.
\]
The variance of the normalized return is
\[
\mathrm{Var}\!\left(\frac{U}{A}\right)
= \frac{1}{A^2}
\left(
\sum_{\alpha} a_\alpha^2 \sigma_\alpha^2
+ 2\sum_{\alpha\neq\beta} a_\alpha a_\beta
\mathrm{Cov}(R_\alpha,R_\beta)
\right).
\]
Diversification effects arise only through the covariance terms, and only when
distinct risk sources are imperfectly correlated.

\paragraph{Cross-sectional diversification (space).}
In portfolio theory, the index set is $\mathcal{I}=\{1,\dots,N\}$, representing
different assets observed over a fixed time interval.
With a fixed budget constraint $\sum_{i=1}^N a_i=A$, diversification reduces
risk when
\[
\mathrm{Corr}(R_i,R_j) < 1 \quad (i\neq j),
\]
so that the covariance terms reduce overall variance.
This is genuine diversification: a fixed exposure is redistributed across
multiple imperfectly correlated risk sources, and the variance parameter of the
resulting return process is reduced.

\paragraph{True temporal diversification (time).}
A directly analogous construction can be made in the time dimension.
Let $\mathcal{I}=\{1,\dots,T\}$ index \emph{non-overlapping} time periods of the
same asset, and suppose that exposure weights $\{a_t\}$ satisfy
\[
\sum_{t=1}^T a_t = A.
\]
If the $R_t$ represent period-specific returns that do not overlap in time,
then
\[
U^{\mathrm{time}} := \sum_{t=1}^T a_t R_t,
\]
and spreading exposure more evenly over time reduces variance under the same
covariance logic as cross-sectional diversification.

An extreme example makes this clear.
Consider two strategies with the same total exposure $A=10$:
\begin{itemize}
\item \emph{Time-concentrated exposure:} invest $10$ units of capital in a
single period and exit immediately.
\item \emph{Time-diversified exposure:} invest $1$ unit of capital in each of
$10$ non-overlapping periods.
\end{itemize}
In the latter case, exposure is genuinely diversified across time, and the
variance of normalized returns is strictly smaller whenever returns across
periods are not perfectly correlated.
This construction is the precise temporal analogue of cross-sectional
diversification.

\paragraph{Why dollar-cost averaging is not temporal diversification.}
Dollar-cost averaging does not satisfy the conditions above.
Capital invested at different times is exposed to the \emph{same} underlying
asset over overlapping periods.
Equivalently, the effective exposure at time $t$ is cumulative,
\[
a_t = \sum_{i\le t} w_i,
\]
so that the pathwise contribution at time $t$ is $a_t R_t$.
As a result, returns associated with capital invested at different times share
overlapping return components and are perfectly positively correlated over those
overlaps.

Because the covariance terms do not decrease, the mechanism responsible for
diversification in either space or true time does not operate under DCA.
Instead, differences between lump-sum and DCA strategies manifest themselves
through differences in exposure profiles and concentration effects.

Related interpretations of DCA as postponing exposure rather than reducing
risk appear in Vanguard (2012) \cite{vanguard2012}.

\paragraph{Risk versus uncertainty.}
In the time-series framework of this paper, the diversification effects
discussed above do not alter risk as defined in Section~2, i.e., the
expectation-level quantity $\mathrm{Risk}(T)=\mathbb{E}[\mathrm{Dev}^2(T)]$.
Rather, they affect the dispersion of realized outcomes across finite samples.
Accordingly, time-diversified exposure reduces \emph{uncertainty}, not risk.

In particular, when uncertainty is compared across strategies, dollar-cost
averaging is more concentrated in the time dimension than either lump-sum
investment or true temporal diversification with uniform exposure.
This explains why DCA may exhibit larger uncertainty than seemingly more
concentrated strategies.

A detailed explanation of why cross-sectional diversification reduces risk
while temporal diversification affects uncertainty in the present setting is
provided in Appendix~D.

\subsection{Lump-sum versus DCA: quantitative risk and uncertainty comparison}

\paragraph{A direct comparison: lump/unit versus DCA (risk and uncertainty).}
We compare DCA to a \emph{uniform temporal exposure} schedule (``lump/unit''),
chosen so that the two strategies have the same return exposure
$E^{(1)}=\sum_{j=1}^t a_j$.

Assume i.i.d.\ Gaussian returns $R_j\sim\mathcal{N}(\mu,\sigma^2)$ and write
$Y_j:=(R_j-\mu)^2$ so that $\mathbb{E}[Y_j]=\sigma^2$ and
$\mathrm{Var}(Y_j)=2\sigma^4$.
Recall that the (pathwise) squared-deviation statistic for a schedule
$\{a_j\}$ is
\[
V_t(\{a_j\}) := \sum_{j=1}^t a_j^2 (R_j-\mu)^2
= \sum_{j=1}^t a_j^2 Y_j .
\]

\emph{DCA.} For evenly spaced DCA with $w_i=1/t$, one has $a_j=j/t$ and
\[
E^{(1)}_{\mathrm{DCA}}=\sum_{j=1}^t \frac{j}{t}=\frac{t+1}{2}.
\]

\emph{Uniform temporal exposure (lump/unit).}
Define $a_j\equiv c$ with the same return exposure as DCA:
\[
\sum_{j=1}^t c = \frac{t+1}{2}
\quad\Longrightarrow\quad
c=\frac{t+1}{2t}.
\]

\smallskip
\noindent\textbf{(i) Risk: $\mathbb{E}[V_t]$.}
By linearity of expectation,
\[
\mathbb{E}[V_t(\{a_j\})]=\sigma^2\sum_{j=1}^t a_j^2.
\]
Hence
\[
\mathbb{E}[V_t^{\mathrm{DCA}}]
=\sigma^2\sum_{j=1}^t \left(\frac{j}{t}\right)^2
=\sigma^2\,\frac{(t+1)(2t+1)}{6t},
\]
while
\[
\mathbb{E}[V_t^{\mathrm{unit}}]
=\sigma^2\sum_{j=1}^t c^2
=\sigma^2\,t\left(\frac{t+1}{2t}\right)^2
=\sigma^2\,\frac{(t+1)^2}{4t}.
\]
Since
\[
\frac{(t+1)(2t+1)}{6t} - \frac{(t+1)^2}{4t}
= \frac{(t+1)(t-1)}{12t} > 0 \qquad (t>1),
\]
we obtain
\[
\mathbb{E}[V_t^{\mathrm{DCA}}] > \mathbb{E}[V_t^{\mathrm{unit}}]
\qquad (t>1),
\]
so DCA is \emph{more time-concentrated} than uniform temporal exposure when
compared at equal return exposure.

\smallskip
\noindent\textbf{(ii) Uncertainty: $\mathrm{Var}(V_t)$.}
Under independence,
\[
\mathrm{Var}(V_t(\{a_j\}))
=\mathrm{Var}\!\left(\sum_{j=1}^t a_j^2 Y_j\right)
= \sum_{j=1}^t a_j^4\,\mathrm{Var}(Y_j)
= 2\sigma^4\sum_{j=1}^t a_j^4.
\]
Therefore
\[
\mathrm{Var}(V_t^{\mathrm{DCA}})
=2\sigma^4\sum_{j=1}^t \left(\frac{j}{t}\right)^4
=2\sigma^4\,
\frac{(t+1)(2t+1)(3t^2+3t-1)}{30\,t^3},
\]
while
\[
\mathrm{Var}(V_t^{\mathrm{unit}})
=2\sigma^4\sum_{j=1}^t c^4
=2\sigma^4\,t\left(\frac{t+1}{2t}\right)^4
=2\sigma^4\,\frac{(t+1)^4}{16\,t^3}.
\]
For $t>1$, one has $\mathrm{Var}(V_t^{\mathrm{DCA}})>\mathrm{Var}(V_t^{\mathrm{unit}})$,
consistent with the fact that non-uniform exposure profiles weaken temporal
averaging and produce larger dispersion of realized risk statistics.

\smallskip
These inequalities formalize the statement that, at equal total return
exposure, DCA is not truly time-diversified: it is more time-concentrated than
uniform temporal exposure both in expected risk and in uncertainty of realized
risk outcomes.

\paragraph{Magnitude of risk differences under timing strategies.}
The comparison above shows that, unlike the long--short horizon case,
lump-sum investment and dollar-cost averaging differ not only in uncertainty but also in expected risk when compared at
equal \emph{return exposure}.
This difference arises because the two strategies induce distinct exposure
profiles over time.

To quantify this effect, consider i.i.d.\ returns with finite fourth moments
and compare dollar-cost averaging to uniform temporal exposure
(``lump/unit'') at equal return exposure
$E^{(1)}=\sum_{j=1}^t a_j=(t+1)/2$.
As shown above,
\[
\mathbb{E}[V_t^{\mathrm{DCA}}]
= \sigma^2\,\frac{(t+1)(2t+1)}{6t},
\qquad
\mathbb{E}[V_t^{\mathrm{unit}}]
= \sigma^2\,\frac{(t+1)^2}{4t}.
\]
Hence
\[
\frac{\mathbb{E}[V_t^{\mathrm{DCA}}]
      -\mathbb{E}[V_t^{\mathrm{unit}}]}
     {\mathbb{E}[V_t^{\mathrm{unit}}]}
= \frac{t-1}{3(t+1)}
\;\longrightarrow\;
\frac{1}{3}
\qquad (t\to\infty).
\]

This shows that dollar-cost averaging is not optimal in the sense of minimizing
expected risk per unit exposure.
However, the deviation from the optimal uniform temporal exposure is of
constant order and does not grow with the investment horizon.
In this sense, DCA is suboptimal but far from pathological.

Importantly, this behavior contrasts sharply with the long--short horizon
comparison in Section~3, where exposure normalization renders expected risk
identical and differences arise solely through uncertainty.
The lump-sum versus DCA comparison therefore provides a concrete counterexample
to the naive intuition that exposure normalization always eliminates
risk differences.

%% file: sections/05_discussion.tex
\section{Discussion: On Model Simplicity and Extensions}

The results presented in Sections~3 and~4 were derived under deliberately
simple probabilistic assumptions.
This simplicity is not a limitation but a methodological feature.
In this section, we discuss why the main conclusions are robust, and clarify
what kinds of additional assumptions would be required to overturn them.

\subsection{On the use of simple models}

A common reaction to results obtained under simple models is to question
their relevance for real markets.
However, simplicity plays a crucial methodological role.

If a claim---such as the assertion that long-term investment or timing
diversification systematically reduces risk---fails to hold even under the
most basic stationary models, then any argument in favor of that claim must
rely on strong additional assumptions.
Such assumptions are necessarily model-dependent and fragile.

In this sense, the burden of proof lies not with the simple model, but with
any proposed mechanism that purports to generate systematic risk reduction.
Absent an explicit mechanism that alters the generating distribution itself,
claims of robust risk reduction cannot be sustained.

\subsection{Negative autocorrelation}

The analysis in Sections~3 and~4 assumed independent returns for clarity.
One might suspect that negative autocorrelation, often associated with mean
reversion, could alter the conclusions.

Negative autocorrelation can indeed reduce the dispersion of cumulative
returns relative to the i.i.d.\ case.
However, this reduction operates through the same channel as before:
it affects aggregation and uncertainty, not the expectation-level definition
of risk adopted in this paper.

In particular, unless autocorrelation is sufficiently strong to induce a
systematic decline in per-period variance over time, cumulative risk continues
to scale with exposure.
The separation between risk and uncertainty therefore remains intact.
A formal treatment of autocorrelated Gaussian returns is provided in
Appendix~B.

\subsection{Heavy-tailed returns}

The results in Sections~3 and~4 rely on the existence of second moments.
When returns have finite variance, the conclusions remain unchanged.

If returns are heavy-tailed with infinite variance, the situation becomes more
extreme rather than more favorable to risk reduction.
In such cases, cumulative risk may grow faster than linearly with exposure,
and uncertainty may fail to concentrate even asymptotically.

Thus, heavy tails do not rescue the notion that long horizons or timing
diversification reduce risk.
If anything, they reinforce the need to distinguish carefully between
expectation-level risk and uncertainty.

\subsection{Nonstationarity}

Finally, consider departures from stationarity.
If the mean $\mu$ or variance $\sigma^2$ of returns changes over time, this
introduces additional uncertainty about future outcomes.
However, such changes do not by themselves constitute a reduction in risk.

For long-term investment or timing diversification to reduce risk in
expectation, the per-period risk itself would have to decline systematically
over time, for example through a structural decrease in $\sigma_t^2$.
Absent such a mechanism, nonstationarity affects uncertainty about parameters
or future realizations, not the accumulation of risk conditional on those
parameters.

This distinction further underscores the central theme of this paper:
apparent reductions in risk are often the result of changes in exposure
profiles or uncertainty, rather than changes in risk itself.

%% file: sections/06_conclusion.tex
\section{Conclusion}

This paper examined two widely held beliefs in investment practice:
that extending the investment horizon reduces risk, and that diversifying
investment timing---most notably through dollar-cost averaging (DCA)---reduces
risk.
By analyzing both issues within a unified probabilistic framework, we showed
that these beliefs originate from related but distinct conceptual
misinterpretations.

When risk is defined consistently at the expectation level as in Section~2,
extending the investment horizon does not alter expected risk or expected
return on a per-unit-exposure basis.
After appropriate exposure normalization, the corresponding risk--return
ratio remains invariant.
Thus, from the standpoint of expectation-level risk or risk--return trade-offs,
the long--short horizon comparison is immaterial.

In contrast, the comparison between lump-sum investment and dollar-cost
averaging exhibits a more subtle structure.
Because these strategies induce different exposure profiles over time,
they may differ not only in uncertainty but also in expected risk when
compared at equal return exposure.
We showed that dollar-cost averaging is not optimal in the sense of minimizing
expected risk per unit exposure, but that the resulting deviation from the
optimal uniform temporal exposure is of constant order and does not grow with
the investment horizon.
In this sense, DCA is suboptimal but far from pathological.

We further showed that both the horizon and timing comparisons generate
differences at the level of uncertainty, defined as the dispersion of realized
outcomes.
Longer horizons lead to greater concentration of realized quantities than
shorter horizons.
Likewise, uniform temporal exposure leads to greater concentration than
time-diversified schedules such as DCA.
These effects, however, concern uncertainty rather than risk itself and should
not be interpreted as general risk reduction mechanisms.

Importantly, the absence of a strong risk-reducing effect does not imply the
reverse conclusion.
The analysis does not support the claim that short-term investment dominates
long-term investment, nor that lump-sum investment uniformly dominates
time-diversified investment.
Once exposure profiles and uncertainty are taken into account, no simple
dominance relation emerges.

From a practical perspective, this implies that choosing long-term investment
or dollar-cost averaging under budgetary, liquidity, or time constraints is
not irrational.
Such choices may be reasonable responses to real-world constraints, even
though they do not, by themselves, guarantee improvements in expectation-level
risk or risk--return characteristics.
At the same time, recognizing the precise scope and limitations of these
mechanisms may help prevent overstated claims of optimality and contribute to
more informed and sustainable investment decision-making.

%% file: sections/99a_appendix.tex
\section{Derivation of the Variance of the Squared Deviation}

In this appendix, we derive the variance of the cumulative squared deviation
\[
V_t := \sum_{i=1}^t (R_i - \mu)^2
\]
and make explicit why a fourth moment condition naturally appears.

Let
\[
Y_i := (R_i-\mu)^2.
\]
Then $V_t = \sum_{i=1}^t Y_i$ and $\mathbb{E}[Y_i]=\sigma^2$ by definition of
$\sigma^2=\mathrm{Var}(R_i)$.

\subsection*{A.1 Why the fourth moment appears}

We begin with the variance of a single squared deviation $Y_1$.
By definition,
\[
\mathrm{Var}(Y_1) = \mathbb{E}\!\left[(Y_1-\mathbb{E}[Y_1])^2\right].
\]
Substituting $Y_1=(R_1-\mu)^2$ and $\mathbb{E}[Y_1]=\sigma^2$ yields
\[
\mathrm{Var}(Y_1)
= \mathbb{E}\!\left[\bigl((R_1-\mu)^2-\sigma^2\bigr)^2\right].
\]
Expanding the square inside the expectation,
\[
\bigl((R_1-\mu)^2-\sigma^2\bigr)^2
= (R_1-\mu)^4 - 2\sigma^2(R_1-\mu)^2 + \sigma^4.
\]
Taking expectations term by term gives
\[
\mathrm{Var}(Y_1)
= \mathbb{E}[(R_1-\mu)^4] - 2\sigma^2\mathbb{E}[(R_1-\mu)^2] + \sigma^4
= \mathbb{E}[(R_1-\mu)^4] - \sigma^4.
\]
Thus, the variance of the squared deviation exists if and only if the fourth
central moment $\mathbb{E}[(R-\mu)^4]$ is finite. This is precisely the reason
that a fourth-moment assumption appears when studying the uncertainty of the
risk statistic $V_t$.

\subsection*{A.2 Variance of the cumulative squared deviation under independence}

Assume that $\{R_i\}$ are independent and identically distributed and that
$\mathbb{E}[(R-\mu)^4] < \infty$. Then $\{Y_i\}$ are i.i.d.\ with
$\mathrm{Var}(Y_i)=\mathrm{Var}(Y_1)$, and
\[
\mathrm{Var}(V_t)
= \mathrm{Var}\!\left(\sum_{i=1}^t Y_i\right)
= \sum_{i=1}^t \mathrm{Var}(Y_i)
= t\,\mathrm{Var}(Y_1).
\]
Consequently,
\[
\mathrm{Var}\!\left(\frac{V_t}{t}\right)
= \frac{\mathrm{Var}(Y_1)}{t}
\;\longrightarrow\; 0
\qquad (t\to\infty),
\]
which justifies the concentration statements in the main text whenever the
fourth moment exists.

\subsection*{A.3 Gaussian specialization: direct computation of the fourth moment}

Suppose $R\sim \mathcal{N}(\mu,\sigma^2)$ and set $Z := (R-\mu)/\sigma$ so that
$Z\sim \mathcal{N}(0,1)$ and $R-\mu=\sigma Z$.
Then
\[
\mathbb{E}[(R-\mu)^4] = \sigma^4 \mathbb{E}[Z^4].
\]
We compute $\mathbb{E}[Z^4]$ directly from the definition:
\[
\mathbb{E}[Z^4]
= \int_{-\infty}^{\infty} z^4 \,\frac{1}{\sqrt{2\pi}}e^{-z^2/2}\,dz.
\]
Using integration by parts with
$u=z^3$ and $dv = z\,\frac{1}{\sqrt{2\pi}}e^{-z^2/2}\,dz$
(so that $du=3z^2dz$ and $v=-\frac{1}{\sqrt{2\pi}}e^{-z^2/2}$), we obtain
\[
\int_{-\infty}^{\infty} z^4 \phi(z)\,dz
= \int_{-\infty}^{\infty} z^3 \bigl(z\phi(z)\bigr)\,dz
= \Bigl[-z^3\phi(z)\Bigr]_{-\infty}^{\infty}
+ 3\int_{-\infty}^{\infty} z^2 \phi(z)\,dz,
\]
where $\phi(z)=\frac{1}{\sqrt{2\pi}}e^{-z^2/2}$ is the standard normal density.
The boundary term vanishes since $z^3\phi(z)\to 0$ as $|z|\to\infty$.
Moreover,
\[
\int_{-\infty}^{\infty} z^2 \phi(z)\,dz = \mathbb{E}[Z^2] = 1.
\]
Hence $\mathbb{E}[Z^4]=3$, and therefore
\[
\mathbb{E}[(R-\mu)^4] = 3\sigma^4.
\]
Substituting into the general expression gives
\[
\mathrm{Var}(Y_1)
= \mathbb{E}[(R-\mu)^4] - \sigma^4
= 2\sigma^4,
\]
and thus
\[
\mathrm{Var}(V_t) = 2t\sigma^4.
\]

This explicit form shows that the uncertainty of the realized per-unit-time
risk statistic $V_t/t$ decays at rate $1/t$ in the Gaussian i.i.d.\ case.

%% file: sections/99b_appendix.tex
\section{Appendix B. Autocorrelated Gaussian Returns}

This appendix extends the analysis of the main text to weakly stationary
Gaussian return processes with autocorrelation.
The purpose is twofold.
First, we show that the expectation of the risk statistic continues to scale
linearly with the horizon (or effective exposure), independently of the
autocorrelation structure.
Second, we show that autocorrelation affects the uncertainty of realized
outcomes, thereby explaining why risk reduction may appear plausible even
though no such reduction occurs in expectation.

\subsection*{B.1 Model setup}

Let $\{R_t\}_{t\ge1}$ be a weakly stationary Gaussian process with
\[
\mathbb{E}[R_t]=\mu,\qquad \mathrm{Var}(R_t)=\sigma^2,
\]
and autocovariance function
\[
\gamma(k):=\mathrm{Cov}(R_t,R_{t+k}),\qquad k\in\mathbb{Z}.
\]
Define the centered process $X_t:=R_t-\mu$. Then
\[
\mathbb{E}[X_t]=0,\qquad \mathbb{E}[X_t^2]=\sigma^2,\qquad
\mathrm{Cov}(X_t,X_{t+k})=\gamma(k).
\]
Autocorrelation modifies joint dependence across time through $\gamma(k)$,
but does not affect the marginal second moment $\mathbb{E}[X_t^2]$ under
stationarity.

\subsection*{B.2 Long-horizon investment: risk and uncertainty}

Recall the cumulative return and risk statistics:
\[
U_t := \sum_{i=1}^t R_i,
\qquad
V_t := \sum_{i=1}^t (R_i-\mu)^2 = \sum_{i=1}^t X_i^2.
\]

\paragraph{Risk (expectation).}
By linearity of expectation,
\[
\mathbb{E}[V_t]
= \sum_{i=1}^t \mathbb{E}[X_i^2]
= t\sigma^2.
\]
This identity holds regardless of the autocovariance structure
$\{\gamma(k)\}$.
Hence, introducing autocorrelation—including negative autocorrelation—does
not alter the linear dependence of expected cumulative risk on the horizon
length $t$.

\paragraph{Uncertainty (dispersion of realized risk).}
To quantify uncertainty, we compute $\mathrm{Var}(V_t)$.
Writing $Y_i:=X_i^2$, we have
\[
\mathrm{Var}(V_t)
= \sum_{i=1}^t \sum_{j=1}^t \mathrm{Cov}(Y_i,Y_j).
\]
Because $(X_i,X_j)$ is jointly Gaussian, Isserlis' theorem yields
\[
\mathrm{Cov}(X_i^2,X_j^2)=2\,\gamma(i-j)^2 \quad (i\neq j),
\qquad
\mathrm{Var}(X_i^2)=2\sigma^4.
\]
Consequently,
\[
\mathrm{Var}(V_t)
= 2t\sigma^4 + 4\sum_{k=1}^{t-1} (t-k)\gamma(k)^2.
\]
Under standard short-memory conditions
(e.g.\ $\sum_k \gamma(k)^2<\infty$), this implies
\[
\mathrm{Var}(V_t)
= t\left(2\sigma^4 + 4\sum_{k=1}^\infty \gamma(k)^2\right) + o(t).
\]
Thus, autocorrelation modifies the uncertainty of realized cumulative risk
through $\gamma(k)^2$, but does not change its linear growth order in $t$.
Negative autocorrelation reduces this uncertainty uniformly across horizons,
but does not generate a long-horizon advantage.

\subsection*{B.3 Time-diversified investment: risk and uncertainty}

We next consider a generic time-diversified investment schedule.
Let $w_i\in[0,1]$ denote the fraction of total capital invested at time $i$,
with $\sum_{i=1}^t w_i=1$, and define the cumulative invested fraction
\[
a_j := \sum_{i=1}^j w_i.
\]

\paragraph{Risk (expectation).}
The contribution of period $j$ to the realized centered return is $a_j X_j$,
so the cumulative risk statistic is
\[
V_t^{\mathrm{TD}} := \sum_{j=1}^t (a_j X_j)^2
= \sum_{j=1}^t a_j^2 X_j^2.
\]
Taking expectations,
\[
\mathbb{E}[V_t^{\mathrm{TD}}]
= \sum_{j=1}^t a_j^2 \mathbb{E}[X_j^2]
= \sigma^2 \sum_{j=1}^t a_j^2.
\]
Once again, autocorrelation does not appear in expectation.
Differences in expected cumulative risk across investment schedules arise
solely from differences in the exposure profile $\{a_j\}$, not from temporal
dependence in returns.

\paragraph{Uncertainty (dispersion of realized risk).}
The uncertainty of the realized risk statistic is given by
\[
\mathrm{Var}(V_t^{\mathrm{TD}})
= \sum_{j=1}^t \sum_{\ell=1}^t a_j^2 a_\ell^2
\,\mathrm{Cov}(X_j^2,X_\ell^2)
= 2\sum_{j=1}^t \sum_{\ell=1}^t a_j^2 a_\ell^2\,\gamma(j-\ell)^2.
\]
Thus, as in the lump-sum case, autocorrelation affects uncertainty through
$\gamma(k)^2$ but does not alter the exposure-driven structure of expected
risk.

\subsection*{B.4 Interpretation}

Autocorrelation alters the dispersion of realized outcomes at all horizons
and for all investment schedules.
In particular, negative autocorrelation reduces the uncertainty of realized
returns and risk statistics symmetrically across short and long horizons.

However, this effect does not constitute a mechanism for risk reduction in
expectation.
The expected cumulative risk continues to scale linearly with the horizon or
effective exposure, both for lump-sum and time-diversified investment.
The apparent safety of long-horizon investment or dollar-cost averaging
under autocorrelation therefore reflects reduced uncertainty, not reduced
risk.

%% file: sections/99c_appendix.tex
\section{A Dice-Roll Example with Population, Realization, and Ensemble Averages}
\label{app:die_example}

This appendix instantiates the hierarchy of quantities used in the main text by an
elementary probability example.
We consider a fair six-sided die rolled $t$ times and define two statistics:
the sum $U_t$ and the cumulative squared deviation $V_t$.
We then compute (i) single-step moments, (ii) population moments of $U_t$ and $V_t$,
(iii) realized values for a concrete sample, and (iv) moments of ensemble averages
over $N$ repeated trials.

\subsection{Setup and notation}

Let $X_1,\dots,X_t$ be i.i.d.\ outcomes of a fair die:
\[
X_i \in \{1,2,3,4,5,6\},
\qquad
\mathbb{P}(X_i=k)=\frac{1}{6}.
\]
Define the single-step mean and variance by
\[
\mu := \mathbb{E}[X_1], \qquad \sigma^2 := \mathrm{Var}(X_1).
\]
Define the $t$-step sum and the cumulative squared deviation (around $\mu$) by
\begin{equation}
U_t := \sum_{i=1}^t X_i,
\qquad
V_t := \sum_{i=1}^t (X_i-\mu)^2.
\label{eq:UtVt_dice}
\end{equation}

For a realized sample $x_1,\dots,x_t$, define the realized counterparts
\begin{equation}
u_t := \sum_{i=1}^t x_i,
\qquad
m_t := \frac{1}{t}\sum_{i=1}^t x_i,
\qquad
v_t := \sum_{i=1}^t (x_i-m_t)^2.
\label{eq:utvt_dice}
\end{equation}
Here $v_t$ is $(t-1)$ times the unbiased sample variance of the realized sample.

\subsection{(1) Single-step mean and variance}

For a fair die,
\begin{equation}
\mu = \mathbb{E}[X_1] = \frac{7}{2},
\qquad
\sigma^2 = \mathrm{Var}(X_1) = \frac{35}{12}.
\label{eq:mu_sigma_dice}
\end{equation}
These correspond to the one-period return moments in the main text.

\subsection{(2) Population moments of $U_t$ and $V_t$}

By independence and linearity of expectation,
\begin{equation}
\mathbb{E}[U_t] = t\mu,
\qquad
\mathrm{Var}(U_t) = t\sigma^2.
\label{eq:Ut_moments}
\end{equation}

For $V_t$, note that $Y_i := (X_i-\mu)^2$ are i.i.d.\ with
$\mathbb{E}[Y_i]=\sigma^2$.
Hence
\begin{equation}
\mathbb{E}[V_t] = t\sigma^2,
\qquad
\mathrm{Var}(V_t) = t\,\mathrm{Var}(Y_1).
\label{eq:Vt_moments_general}
\end{equation}

In the fair-die case, one can compute
\[
\mathbb{E}[Y_1^2]
= \mathbb{E}\!\left[(X_1-\mu)^4\right]
= \frac{707}{48},
\qquad
\mathrm{Var}(Y_1)
= \mathbb{E}[Y_1^2]-\big(\mathbb{E}[Y_1]\big)^2
= \frac{56}{9},
\]
and therefore
\begin{equation}
\mathrm{Var}(V_t) = t\cdot\frac{56}{9}.
\label{eq:VarVt_dice}
\end{equation}

Specializing to $t=6$, we obtain
\begin{equation}
\mathbb{E}[U_6]=6\mu=21,
\qquad
\mathrm{Var}(U_6)=6\sigma^2=\frac{35}{2},
\label{eq:Ut_t6}
\end{equation}
\begin{equation}
\mathbb{E}[V_6]=6\sigma^2=\frac{35}{2},
\qquad
\mathrm{Var}(V_6)=6\cdot\frac{56}{9}=\frac{112}{3}.
\label{eq:Vt_t6}
\end{equation}

\subsection{(3) A concrete realized sample for $t=6$}

Consider the realized outcomes
\[
x_1, x_2, x_3, x_4, x_5, x_6 = 1,\,5,\,6,\,6,\,4,\,2.
\]
Then, by direct calculation,
\begin{equation}
u_6 = 24,
\qquad
m_6 = \frac{u_6}{6} = 4,
\qquad
v_6 = \sum_{i=1}^6 (x_i-m_6)^2 = 22.
\label{eq:realized_t6}
\end{equation}
These correspond to one realized $t$-period return and risk statistics in the main text.

\subsection{(4) Ensemble averages over $N$ repeated $t$-step trials}

To model the ``many possible paths'' viewpoint, consider repeating the $t$-roll
experiment $N$ times independently.
Let $\{U_t^{(j)},V_t^{(j)}\}_{j=1}^N$ be i.i.d.\ copies of $(U_t,V_t)$.
Define the ensemble (cross-trial) averages
\begin{equation}
\bar U_{t,N} := \frac{1}{N}\sum_{j=1}^N U_t^{(j)},
\qquad
\bar V_{t,N} := \frac{1}{N}\sum_{j=1}^N V_t^{(j)}.
\label{eq:ensemble_averages}
\end{equation}

By standard properties of i.i.d.\ averages,
\begin{equation}
\mathbb{E}[\bar U_{t,N}] = \mathbb{E}[U_t] = t\mu,
\qquad
\mathrm{Var}(\bar U_{t,N}) = \frac{\mathrm{Var}(U_t)}{N}
= \frac{t\sigma^2}{N},
\label{eq:ensemble_U}
\end{equation}
and
\begin{equation}
\mathbb{E}[\bar V_{t,N}] = \mathbb{E}[V_t] = t\sigma^2,
\qquad
\mathrm{Var}(\bar V_{t,N}) = \frac{\mathrm{Var}(V_t)}{N}
= \frac{t\,\mathrm{Var}(Y_1)}{N}.
\label{eq:ensemble_V}
\end{equation}

Specializing to $t=6$, we have
\begin{equation}
\mathbb{E}[\bar U_{6,N}] = 21,
\qquad
\mathrm{Var}(\bar U_{6,N}) = \frac{35/2}{N},
\label{eq:ensemble_U_t6}
\end{equation}
\begin{equation}
\mathbb{E}[\bar V_{6,N}] = \frac{35}{2},
\qquad
\mathrm{Var}(\bar V_{6,N}) = \frac{112/3}{N}.
\label{eq:ensemble_V_t6}
\end{equation}

In particular, for $N=1$ these reduce to the population moments of $(U_6,V_6)$,
while as $N\to\infty$ the variances vanish:
\[
\mathrm{Var}(\bar U_{6,N})\to 0,
\qquad
\mathrm{Var}(\bar V_{6,N})\to 0.
\]
This illustrates how ``more paths'' (larger $N$) yields more concentrated
ensemble averages, without changing the underlying data-generating process.

%% file: sections/99d_appendix.tex
\section{Risk, Uncertainty, and Variance Decomposition}

This appendix clarifies the relationship between variance decomposition in
cross-sectional portfolio theory and the separation of risk and uncertainty
adopted in the present time-series framework.
Although the algebraic expressions may appear similar, the objects being
decomposed differ fundamentally.

\subsection{Cross-sectional variance decomposition}

In cross-sectional portfolio theory, one considers a vector of single-period
returns $R=(R_1,\dots,R_N)$ with covariance matrix
\[
\Sigma = \mathbb{E}\bigl[(R-\mu)(R-\mu)^\top\bigr],
\]
which is a parameter of the data-generating process.
For portfolio weights $w$ satisfying $\sum_i w_i=1$, the portfolio return
$R^{\mathrm{port}}=w^\top R$ has variance
\[
\mathrm{Var}(R^{\mathrm{port}})=w^\top\Sigma w.
\]
This quantity is itself a population-level variance parameter.
Standard variance decompositions separate this into a weighted volatility term
and a diversification effect, but both components pertain to the variance
parameter of the newly constructed return process.
In particular, the resulting portfolio volatility is a property of the
generating distribution, not of any realized sample.

\subsection{Time-series risk and realized volatility}

In contrast, the time-series analysis in this paper begins with pathwise
quantities.
For a single realized return path $\{r_t\}_{t=1}^T$, the realized volatility
\[
\hat\sigma_T^2 := \frac{1}{T}\sum_{t=1}^T (r_t-\bar r)^2
\]
is itself a random variable.
Expectation-level risk is defined instead as
\[
\mathrm{Risk}(T):=\mathbb{E}\!\left[\sum_{t=1}^T (R_t-\mu)^2\right]
= T\sigma^2,
\]
which is a parameter of the generating distribution and does not depend on the
particular realization.

Because $\hat\sigma_T^2$ is random, it cannot be reconstructed from
$\mathrm{Risk}(T)$ and $\mathrm{Var}(\mathrm{Dev}^2(T))$ alone.
Finite-sample realized volatility therefore lies outside the scope of
expectation-level variance decomposition.

\subsection{Why the analogy breaks}

The apparent analogy between
$\mathrm{Var}(w^\top R)$ in the cross-sectional setting and
$\mathrm{Var}(\sum_{t=1}^T R_t)$ in the time-series setting is misleading.
In the former case, the variance pertains to a newly defined random variable
and describes a property of its generating distribution.
In the latter case, the variance quantifies dispersion of realized outcomes
around an unchanged generating distribution.

Thus, while cross-sectional diversification reduces risk by constructing a
new return process with a smaller variance parameter, temporal aggregation
does not alter the variance parameter of the underlying return process.
Instead, it affects only the uncertainty associated with finite-sample
realizations.

\subsection{Summary}

The distinction can be summarized as follows:
cross-sectional variance decompositions recover population-level volatility,
whereas time-series variance calculations concern realized volatility.
The separation of risk and uncertainty adopted in this paper is therefore
essential for maintaining conceptual consistency in finite-horizon analysis.